\documentclass[sigconf]{acmart}

\usepackage{graphicx}
\usepackage{epsfig}
\usepackage{subfig}
\usepackage{booktabs}
\usepackage{amsmath}
\usepackage{amssymb}

\newcommand{\ie}{\textit{i}.\textit{e}.}
\newcommand{\eg}{\textit{e}.\textit{g}.}

\def\BibTeX{{\rm B\kern-.05em{\sc i\kern-.025em b}\kern-.08emT\kern-.1667em\lower.7ex\hbox{E}\kern-.125emX}}
    
\begin{document}

%
% The "title" command has an optional parameter, allowing the author to define a "short title" to be used in page headers.
\title{Intrinsic Image Popularity Assessment}

%
% The "author" command and its associated commands are used to define the authors and their affiliations.
% Of note is the shared affiliation of the first two authors, and the "authornote" and "authornotemark" commands
% used to denote shared contribution to the research.

\author{Keyan Ding}
\affiliation{%
  \institution{City University of Hong Kong}
  \city{Hong Kong}
}
\email{keyanding2-c@my.cityu.edu.hk}

\author{Kede Ma}
\affiliation{%
  \institution{City University of Hong Kong}
  \city{Hong Kong}
  %\country{USA}
  }
\email{kede.ma@cityu.edu.hk}

\author{Shiqi Wang}
\affiliation{%
  \institution{City University of Hong Kong}
  \city{Hong Kong}
}
\email{shiqwang@cityu.edu.hk}

%
% By default, the full list of authors will be used in the page headers. Often, this list is too long, and will overlap
% other information printed in the page headers. This command allows the author to define a more concise list
% of authors' names for this purpose.

%
% The abstract is a short summary of the work to be presented in the article.
\begin{abstract}
   The goal of research in automatic image popularity assessment (IPA) is to develop computational models that can accurately predict the potential of a social image to go viral on the Internet. Here, we aim to single out the contribution of visual content to image popularity, \ie, intrinsic image popularity. Specifically, we first describe a probabilistic method to generate massive popularity-discriminable image pairs, based on which the first large-scale image database for intrinsic IPA (I$^2$PA) is established.  We then develop computational models for I$^2$PA based on deep neural networks, optimizing for ranking consistency with  millions of popularity-discriminable image pairs. Experiments on Instagram and other social platforms demonstrate that the optimized model performs favorably against existing methods, exhibits reasonable generalizability on different databases, and even surpasses human-level performance on Instagram. In addition, we conduct a psychophysical experiment to analyze various aspects of human behavior in I$^2$PA.
   %and find that our model slightly surpasses the human-level performance on Instagram.  
\end{abstract}

\begin{CCSXML}
<ccs2012>
<concept>
<concept_id>10002951.10003317.10003318.10003321</concept_id>
<concept_desc>Information systems~Content analysis and feature selection</concept_desc>
<concept_significance>500</concept_significance>
</concept>
<concept>
<concept_id>10003120.10003130.10003131.10011761</concept_id>
<concept_desc>Human-centered computing~Social media</concept_desc>
<concept_significance>500</concept_significance>
</concept>
% <concept_id>10010147.10010257.10010258.10010259.10003343</concept_id>
% <concept_desc>Computing methodologies~Learning to rank</concept_desc>
</ccs2012>
\end{CCSXML}

\ccsdesc[500]{Information systems~Content analysis and feature selection}
\ccsdesc[500]{Human-centered computing~Social media}
% \ccsdesc[500]{Computing methodologies~Learning to rank}

%
% Keywords. The author(s) should pick words that accurately describe the work being presented. Separate the keywords with commas.
\keywords{Intrinsic image popularity, learning-to-rank, deep neural networks, human behavior analysis.}

%
% This command processes the author and affiliation and title information and builds the first part of the formatted document.
\maketitle

%----------------------------------------------------------------------
\section{Introduction}
Recent years have witnessed an accelerated proliferation of images and videos being uploaded to various social platforms such as Instagram\footnote{\url{https://www.instagram.com}}, Flickr\footnote{\url{https://www.flickr.com}}, and Reddit\footnote{\url{https://www.reddit.com}}. Some photos turn to be extremely popular, which gain millions of likes and comments, while some are completely ignored. Even for images uploaded by the same user at the same time, their popularity may be substantially different. This interesting phenomenon motivates us to ask what the secret of image popularity is. It is generally believed that image popularity is determined by a combination of factors, including the visual content, the user statistics, the upload time, and the caption~\cite{pinto2013using,gelli2015image,khosla2014makes,mcparlane2014nobody}. 

\begin{figure}[t]
\centering
  \includegraphics[width=1\linewidth]{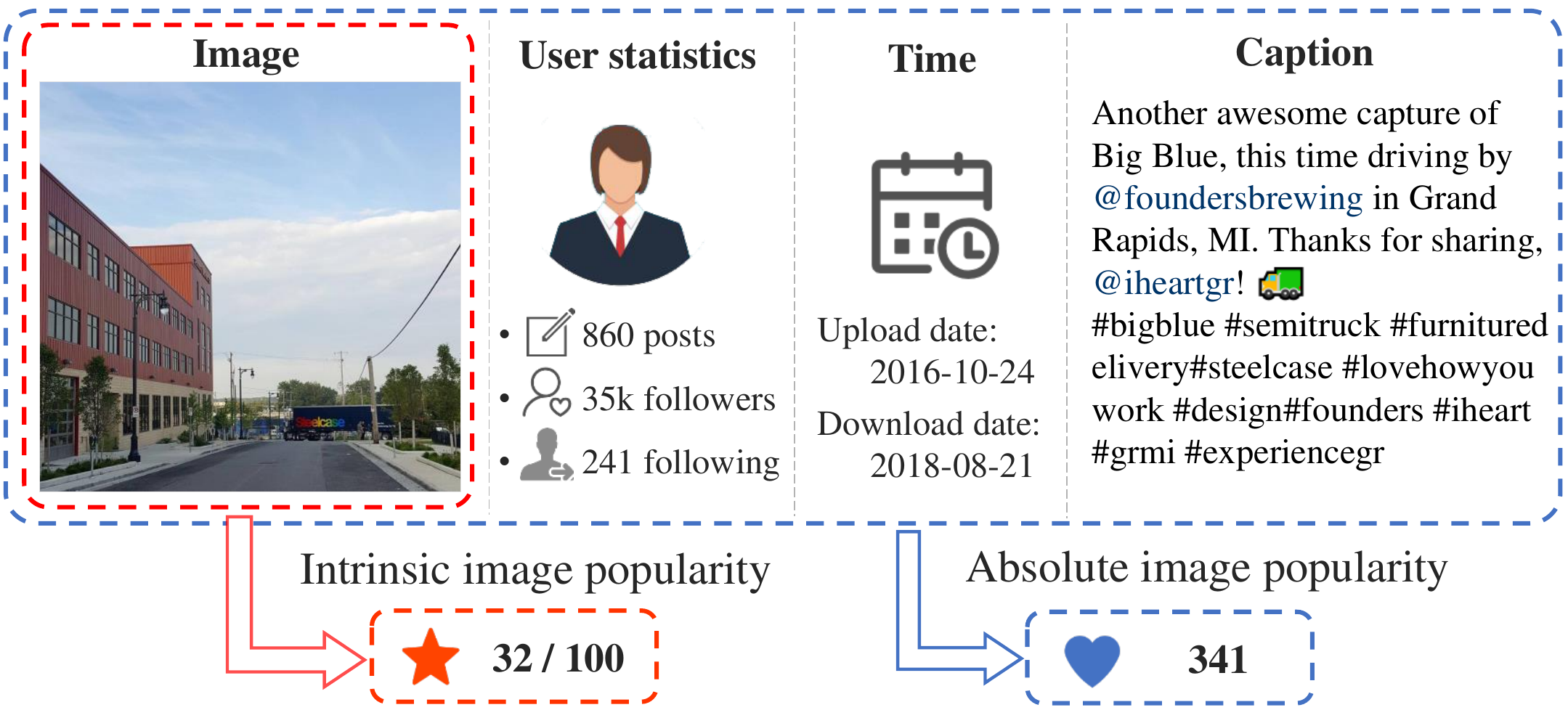}
  \caption{Absolute IPA versus I$^2$PA. Absolute IPA makes use of the image and all its relevant social and textual information to predict the number of received likes, while I$^2$PA relies solely on the image itself, exploiting the visual content for popularity prediction. 341 is the number of likes received by the image, and 32 is the predicted intrinsic popularity score by our model (with a re-scaled maximum score of 100).}
  \label{fig:demo}
  \vspace{-3mm}
\end{figure}

Computational models for {\em absolute} image popularity assessment (IPA) attempt to predict the number of received likes/comments of an image by combining all visual and non-visual factors~\cite{khosla2014makes,mcparlane2014nobody}. Here we aim to single out the contribution of visual content to image popularity, namely {\em intrinsic} image popularity, and develop computational models for intrinsic IPA (I$^2$PA) for several reasons. First, by focusing on the visual content, I$^2$PA is a cleaner and easier-to-interpret problem than absolute IPA (see Fig.~\ref{fig:demo}).
%{\em absolute} image popularity assessment (IPA) (\eg, predicting the number of received likes) is difficult because of the complicated interactions among all relevant factors.  By contrast, intrinsic IPA (I$^2$PA), which focuses on the contribution of the visual content, is a cleaner problem .  
Second, computational models for I$^2$PA guide the identification of potentially popular images with no social and textual contexts, and hold much promise in optimizing social image management and recommendation systems in the long run. For example, more computation and storage resources may be allocated to images with high intrinsic popularity. Third, from the users' perspective, I$^2$PA model predictions are ideal indicators of which images in their personal albums are worth uploading to gain great attention, when they just join the social network and have no social interactions. 
Moreover, the users may gain inspiration regarding how to filter and prioritize photos assisted by the model instead of their own biased opinions. Last, analyzing how image attributes such as image quality, aesthetics, contexts, and semantics contribute to intrinsic image popularity is by itself an interesting problem for human and computer vision study (see Sections~\ref{subsec:quan} and~\ref{subsec:qual}).

In this paper, we conduct a systematic study of I$^2$PA based on Instagram, a leading photo-sharing social network with over one billion monthly active users on the web and mobile clients~\cite{1bi}. %Instagram emphasizes the self-expression of images and characterizes them through texts and tags. 
% {\em Absolute} image popularity can be quantified in a variety of ways~\cite{khosla2014makes} and we treat the number of likes received by a photo as its ground truth, which is ubiquitous on Instagram. 
We first develop a probabilistic method to construct a new form of data - popularity-discriminable image pairs (PDIPs), which contain rich information about intrinsic image popularity by reducing the influence of non-visual factors. We show that such PDIPs can be generated at very low cost and high accuracy, leading to the first large-scale image database for I$^2$PA.
% with the help of {\em absolute} image popularity scores. 
We then train deep neural networks (DNNs) to predict intrinsic image popularity by  learning-to-rank~\cite{burges2005learning,chopra2005learning,joachims2002optimizing} millions of PDIPs in the proposed database.
Experimental results on Instagram and several other social platforms show that our  model  predicts intrinsic image popularity accurately, outperforms state-of-the-art methods (\eg, the commercialized products Virality Detection~\cite{virality} and LikelyAI~\cite{likelyai}), and generalizes reasonably. Moreover, we conduct a psychophysical experiment to collect human opinions on intrinsic image popularity, and find that our method slightly surpasses the human-level performance on Instagram.  

Our contributions are four-fold. First, we revisit the concept of intrinsic image popularity with a new problem formulation. Second, we construct the first large-scale image database for I$^2$PA, consisting of more than two million PDIPs with reliable annotations. Third, we develop a computational model for I$^2$PA based on a DNN, which delivers human-level performance. Fourth, we conduct a psychophysical experiment to analyze various aspects of human behavior in I$^2$PA.

%----------------------------------------------------------------
\section{Related Work}
Popularity assessment of social media content (\eg, texts, photos, and videos) has been an active research field in the past decade. Traditional computer vision and natural language processing methods focused on handcrafting image and text features~\cite{damashek1995gauging,lowe2004distinctive}, which requires extensive human expertise.  Recently, there has been a roaring wave of developing DNNs that emphasize automatic hierarchical feature learning for IPA~\cite{hsu2017social,trzcinski2017predicting,likelyai,virality}. 

Most studies investigate image popularity based on the professional photo-sharing site Flickr. McParlane \textit{et al.}~\cite{mcparlane2014nobody} predicted image popularity (\ie, the number of comments and views) using image and user contexts. Khosla \textit{et al.}~\cite{khosla2014makes} predicted the normalized view counts, and analyzed the impact of low-level features (color patch variance, gradient, and texture), middle-level features (GIST~\cite{oliva2001modeling}) and high-level features (semantics) on the prediction accuracy. Gelli \textit{et al.}~\cite{gelli2015image} conducted a qualitative analysis regarding which visual factors may influence image popularity. Wu \textit{et al.}~\cite{wu2016time,Wu2017DTCN,wu2016unfolding} incorporated multiple time-scale dynamics in predicting image popularity. Zhang \textit{et al.}~\cite{Zhang2018User} proposed a user-guided hierarchical attention network for multi-modal content popularity prediction.

% Instagram is another social network to study image popularity. Mazloom \textit{et al.}~\cite{mazloom2016multimodal} examined several engagement parameters, including factual information, sentiment, vividness, and entertainment to predict the popularity of brand-related posts. They later predicted content-based post popularity using various visual and textual features~\cite{mazloom2018category}.  Almgren \textit{et al.}~\cite{almgren2016predicting} attempted to predict future image popularity by employing the social contexts, the semantics, and the early popularity. 

There are also several studies of IPA using other social networks. Mazloom \textit{et al.}~\cite{mazloom2016multimodal} 
%examined several engagement parameters, including factual information, sentiment, vividness, and entertainment to 
predicted the popularity of brand-related posts on Instagram, and  later extended their method to account for category specific posts~\cite{mazloom2018category}.  %Almgren \textit{et al.}~\cite{almgren2016predicting} assessed future image popularity on Instagram. %by employing the social contexts, the semantics, and the early popularity.
Zhang \textit{et al.}~\cite{zhang2018become} addressed user-specific popularity prediction on Instagram with a dual-attention mechanism.
Deza and Parikh~\cite{deza2015understanding} cast IPA as a classification problem based on images collected from Reddit, a popular website composed of many interest-centric sub-communities. Hessel \textit{et al.}~\cite{hessel2017cats} compared multi-modal content with social contexts in predicting relative popularity on Reddit. They found that visual and textual features tend to outperform user statistics.

Due to its commercial value, many companies have  developed computational models for IPA. LikelyAI~\cite{likelyai} is such a product that assesses the  popularity of Instagram posts. Trained on millions of images, LikelyAI is claimed to recognize popular patterns on Instagram.  Virality Detection~\cite{virality} is a similar tool to score images based on their potentials to become popular on social media. Virality Detection is trained on a massive corpus of web images, and achieves high accuracy on the AVA dataset~\cite{murray2012ava}, suggesting close agreement between image aesthetics and image popularity.

Most of the above-mentioned methods try to predict absolute image popularity by combining various visual and non-visual features.  In contrast, I$^2$PA receives little attention despite broad practical applications. The closest studies to ours are due to Cappallo \textit{et al.}~\cite{cappallo2015latent}, and Dubey and Agarwal~\cite{dubey2017modeling}. Cappallo \textit{et al.} learned a visual popularity predictor from both popular and unpopular images on Flickr using RankSVM~\cite{joachims2002optimizing}. However, their training pair generation process does not exclude the impact of non-visual factors such as user statistics and textual information. Moreover, their model is not end-to-end optimized, and may result in suboptimal performance. Dubey and Agarwal~\cite{dubey2017modeling} modeled image popularity with pairwise spatial transformer networks, whose training pairs suffer from similar problems~\cite{cappallo2015latent}. In addition, the image content based on Reddit ~\cite{deza2015understanding} is not diverse enough, and may hinder the generalizability of the learned networks.

% There are some other products such as CorneaAI\footnote{\url{https://betalist.com/startups/cornea-ai}}, Beautiful Destinations\footnote{\url{https://www.beautifuldestinations.com}}, can predict the popularity of photos to help clients enhance their photo sharing experience.

%=====================================================================

\section{Method}
In this section, we first describe the probabilistic method for PDIP generation and ways to reduce the impact of non-visual factors. 
%with emphasis on the user statistics, the upload time, and the caption. 
Based on the method, we build the first large-scale image database for I$^2$PA. Last, we describe the specification and learning of our DNN-based computational model for I$^2$PA.

\subsection{PDIP generation and database construction}\label{pdip_gen}
At the beginning, we crawl more than one million active users on Instagram using the snowball sampling~\cite{biernacki1981snowball} through the chain of followers. To decorrelate the sampled users, we randomly remove $80\%$ of them, and collect the information of each post of the remaining users, including the download time, the post URL, the user ID, the content type, the upload time, the caption (including emojis, hashtags, and @ signs), the number of likes, and the number of comments. As a result, we obtain over $200$ million distinctive posts as the candidates to build our database. Note that all data are collected via HTTP requests for research purpose only.
 
We present in detail the probabilistic method for PDIP generation. In agreement with~\cite{almgren2016predicting,mazloom2018category,mazloom2016multimodal}, the log-scaled number of likes $S$ is considered as the ground truth for absolute image popularity, based on which we make two mild assumptions.
\begin{itemize}
\item $S$ obeys a normal distribution (assuming the Thurstone's model~\cite{thurstone1927law})
\begin{align}\label{eq:lognormal}
p(S\vert \mu) \propto \exp \left(-\frac{\left( S -\mu\right)^2}{2\sigma^2}\right)\,,
\end{align}
%where $\ln(\cdot)$ stands for the natural logarithmic function and is often used~\cite{khosla2014makes,wu2016time,Wu2017DTCN} to pre-process the number of likes $S$. 
with mean $\mu$ and standard deviation (std) $\sigma$. Here $\mu$ is a random variable, which can be viewed as the average number of likes received by an image in the log scale, if the image were uploaded and rated multiple times.  Without any prior knowledge, we assume $p(\mu)$ is flat with a finite positive support. To simplify the derivation, we treat $\sigma$ as a positive constant to be determined.  
\item The intrinsic image popularity $Q$ is a monotonically increasing function of $\mu$. 
\end{itemize} 
Using the Bayes' theorem, we have
\begin{align}\label{eq:bt}
p(\mu\vert S)\propto p(S\vert\mu)p(\mu)\propto p(S\vert\mu)\,,
\end{align}
where the second proportion follows from the assumption that $p(\mu)$ is flat. That is, conditioning on $S$, $\mu$ is Gaussian with  mean $S$ and  std $\sigma$. \begin{table}[t]
  \centering
  \caption{Statistics of the proposed large-scale image database for I$^2$PA}
    \begin{tabular}{l|l}
    \toprule
    Attribute & Value \\ 
    \hline
    Number of PDIPs & $2.5 \times 10^6$ \\
    Number of users involved & $1.1 \times 10^5$ \\
    Average likes per image & $5.3 \times 10^3$ \\
    Average $P(Q_A\ge Q_B\vert S_A,S_B)$ & $0.978$ \\
    Average upload time interval & $4.8$ days \\
    Proportion of no hashtag & $45.6\%$ \\
    Proportion of no @ sign & $47.9\%$ \\
    Proportion of no caption & $11.1\%$\\
    Average length of descriptive text & $2.1$ words \\ 
    \bottomrule
    \end{tabular}
  \label{tab:pdips}
\end{table}
To ensure that Image $A$ is intrinsically more popular than Image $B$ in a PDIP, we compute the probability  
\begin{align}
\label{eq:p} P(Q_A\ge Q_B\vert S_A, S_B)=&P(\mu_A\ge \mu_B\vert S_A, S_B)\\
=&P(\mu_A-\mu_B \ge 0\vert S_A, S_B)\,,
\end{align}
where Eq.~(\ref{eq:p}) follows from the assumption that $Q$ is a monotonically increasing function of $\mu$.   Assuming  the  variability  of intrinsic popularity across images is uncorrelated, and conditioning on $S_A$ and $S_B$, the difference $\mu_{AB}=\mu_A - 
\mu_B$  is also Gaussian
\begin{align}\label{eq:normaldiff}
p(\mu_{AB}\vert S_A, S_B) \propto \exp \left(-\frac{\left(
\mu_{AB} - \left(S_A- S_B\right)\right)^2}{4\sigma^2}\right)\,.
\end{align}
Combining Eq.~(\ref{eq:p}) with Eq.~(\ref{eq:normaldiff}),  we have 
\begin{align}\label{eq:p2}
P(Q_A \ge Q_B\vert S_A, S_B)=\Phi\left(\frac{S_A -  S_B}{\sqrt{2}\sigma}\right)\,,
\end{align}
where $\Phi(\cdot)$ is the standard normal cumulative distribution function. $P(Q_A\ge Q_B\vert S_A, S_B)$ indicates the probability that Image \textit{A} is intrinsically more popular than Image \textit{B}. In practice, we choose a large threshold $P(Q_A\ge Q_B\vert S_A, S_B)\ge T$ to ensure the popularity discriminability of PDIPs.

%--------six samples pairs-----
\begin{figure} 
    \centering
    \includegraphics[width=1\linewidth]{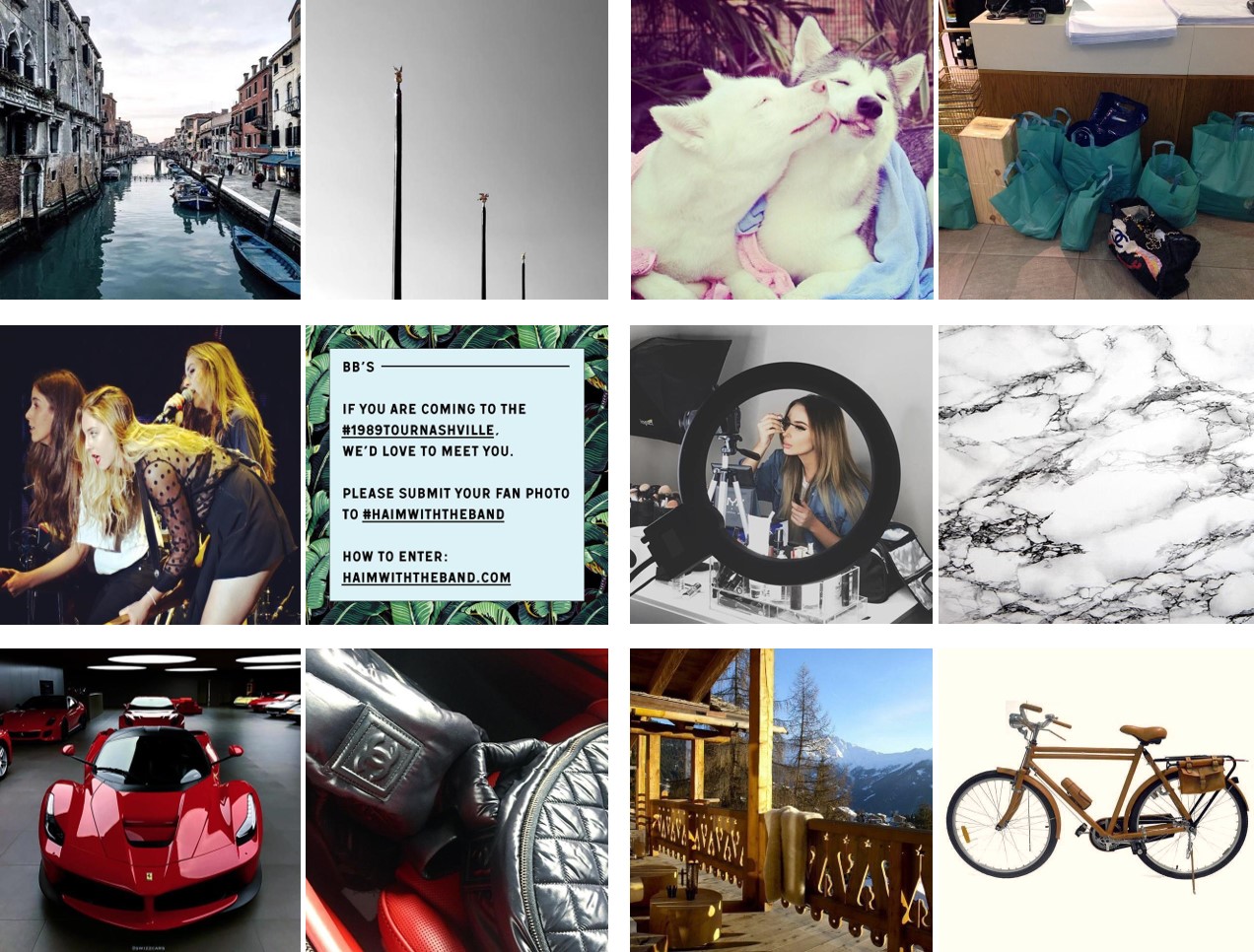}
	  \caption{Representative PDIPs from the proposed database. The left image in each pair is expected to be intrinsically more popular than the right one, which has been confirmed by our psychophysical experiment. More than 80\% of the subjects think the left images would receive more likes on social media than the right ones.}
	  \label{fig:spdip} 
% 	  \vspace{-3mm}
\end{figure}

Enumerating all possible pairs that only satisfy the probability constraint is not enough, as image popularity may be affected by other non-visual factors. Therefore, it is desirable to further constrain the two images in a pair to have similar textual and social contexts. 
% in order to single out the contribution of the visual content to image popularity, \ie, the desired {\em intrinsic} image popularity. 
According to the mechanism of Instagram, we consider three major non-visual factors.

\begin{itemize} 
\item \textbf{User statistics.} Several studies have showed that the popularity of an image is highly correlated with the user who uploads it~\cite{can2013predicting,gelli2015image,yamaguchi2014chic}. The most obvious reason is that different users have different numbers of followers. Images posted by the users with more followers have higher chances of receiving more likes. The number of active followers and their preferences make the relationship more complicated. Considering the above issues, we restrict images from the same user for PDIP generation.
\item \textbf{Upload time.} A user often has a different number of followers at different times. To reduce this effect,  the post time difference of the two images in a PDIP is set to a maximum of ten days. In addition, it is helpful to exclude images just uploaded to the social network as the number of likes has not reached a stable state. According to the analysis in \cite{almgren2016predicting}, the number of likes for most images stops to increase after four weeks. As such, we exclude images posted within one month. In addition, the upload time in a day may also affect image popularity. Failing to model this issue may result in minor label noise. However, as will be clear in Section~\ref{subsec:quan}, our learning process  is quite robust to label noise in PDIPs.
\item \textbf{Caption.} Image captions have a noticeable impact on  image popularity, especially those containing hashtags and @ signs. A hot hashtag contributes significantly to image popularity because of the extensive exposure to viewers beyond followers. Generally the more hashtags of a post, the greater chances of receiving more likes. @ signs may also affect image popularity. For example, images @ a celebrity would probably receive more likes than those @ an ordinary user or without the @ sign. To remove the textual bias, we restrict the hashtag and @ sign of the images in a PDIP to be the same (in both content and number). Moreover, the length of the caption (excluding the hashtag and @ sign) is restricted to a maximum of six words.
\end{itemize}

We summarize the four constraints of images for PDIP generation as follows:
% \begin{enumerate}
%  \setlength{\itemsep}{1pt}
%  \setlength{\parskip}{0pt}
%  \setlength{\parsep}{0pt}
% \item Two images come from the same user;
% \item Two images have been posted over one month, whose upload time interval is less than $10$ days;
% \item Two images have less than $6$ words of descriptive text with the same hashtags and @ a person;
% \item The threshold $T$ for the probability of A over B $P(Q_A > Q_B)$ in Eq.~(\ref{eq:p2}) is set to $0.9$.
% \end{enumerate}
\begin{itemize}
\setlength{\itemsep}{1pt}
\setlength{\parskip}{0pt}
\setlength{\parsep}{0pt}
\item $P(Q_A\ge Q_B\vert S_A, S_B)\ge T$;
\item from the same user;
\item posted more than one month and within ten days;
\item caption with a maximum of six words and the same hashtag/@ sign.
\end{itemize}
An Instagram post may contain multiple images, videos, or a mixture of them. Here we only consider single image posts  because it is difficult to allocate the number of likes across multiple images in a post. In addition, we exclude images with less than $50$ likes to reduce the boundary effect. From more than $200$ million candidate images on Instagram, we construct the first large-scale database for I$^2$PA, which contains approximately $2.5$ million of PDIPs, satisfying all of the above constraints. To ensure the content diversity, one image only participates in one PDIP.  Table~\ref{tab:pdips} summarizes the statistics of the proposed database, and  Fig.~\ref{fig:spdip} shows six sample PDIPs.
%, where our PDIP generation engine considers the left images to be clearly more popular than the right ones, which is confirmed by our psychophysical experiment (Section~\ref{subsec:quan}). 

% Figure 3 seems a little wrong. The two intrinsic popularity scores should not directly connect to the cross entropy loss. According to the paper, the difference between the two scores is calculated, and then passed through a logistic function before the loss is calculated.

%------------------------ proposed model ----------------------------

\subsection{DNN-based computational models for I$^2$PA}
\begin{figure}
\centering
  \includegraphics[width=1\linewidth]{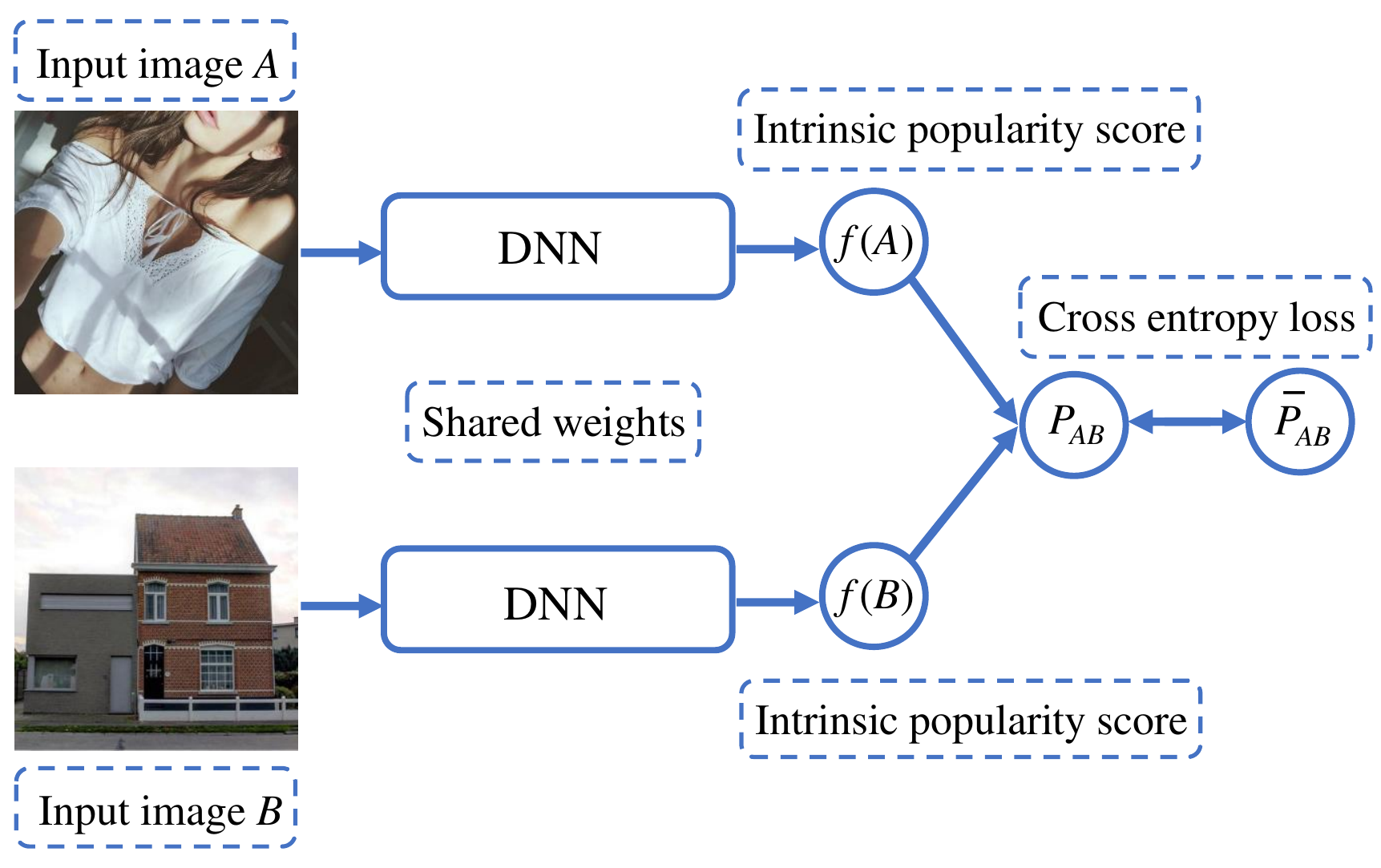}
  \caption{Computational models for I$^2$PA based on DNNs. The two streams have the same network architecture, whose model parameters are shared and optimized by minimizing the binary cross entropy loss. Either stream can be used to predict intrinsic image popularity.}
  \label{fig:architecture}
  \vspace{-3mm}
\end{figure}

In this subsection, we describe a DNN-based computational model for I$^2$PA by learning-to-rank millions of PDIPs. As a machine learning technique, learning-to-rank was extensively studied in the context of information retrieval~\cite{liu2009learning}, and later found its way to computer vision~\cite{chopra2005learning}, image processing~\cite{ma2017dipiq}, and natural language processing~\cite{li2011learning}. Pairwise learning-to-rank approaches assume that the relative order between two instances is known (or can be inferred), and aim to minimize the average number of incorrectly ordered pairs. The PDIPs in our database fit the pairwise learning-to-rank scheme naturally, and we use them  to drive the learning of a Siamese architecture for I$^2$PA (see Fig.~\ref{fig:architecture}). 

The input $A$ of a PDIP to the first stream is an RGB image, and the output is  the predicted intrinsic popularity score $Q_A = f(A)$. Similarly, the second stream inputs the other image $B$ and predicts $Q_B=f(B)$. The network architectures of the two streams are the same, whose weights are shared during training and testing. We compute the predicted score difference $O_{AB} = f(A) - f(B)$, and convert it to a probability using a logistic function
\begin{align}
{P_{AB}} = \frac{\exp\left(O_{AB}\right)}{1 + \exp\left(O_{AB}\right)}\,.
\end{align}
Denote the ground-truth binary label of a PDIP as $\bar{P}_{AB}$, where $\bar{P}_{AB} = 1$ indicates Image $A$ is intrinsically more popular than Image $B$ and otherwise $\bar{P}_{AB} = 0$. We adopt the binary cross entropy as the loss function
\begin{align}
\ell =& -\bar{P}_{AB}\log P_{AB}  - \left(1-\bar{P}_{AB}\right)\log\left(1-P_{AB}\right)\\
=&-\bar{P}_{AB}O_{AB} + \ln\left(1 + \exp\left(O_{AB}\right)\right)\,.
\end{align}

After training, the optimal predictor $f^\star$ (either stream in the Siamese architecture) is learned. Given a test image $X$, we perform a standard forward pass to obtain the predicted  intrinsic popularity score
\begin{align}
Q_X = f^\star(X)\,.
\end{align}

%===========================================================================

\section{Experiments}
In this section, we first describe the implementation details, including the default DNN architecture and the training procedure. We then quantitatively compare our model with the state-of-the-art. %We further show the generalizability of our method by applying it to other social platforms. 
We also conduct qualitative analysis of our model, and have a number of interesting observations. Last, we perform a psychophysical experiment to analyze human behavior in this task.

\subsection{Implementation details}\label{subsec:id}
 We adopt ResNet-50~\cite{he2016deep} as our default DNN architecture, and replace the last layer with a fully connected layer of one output, representing the predicted intrinsic popularity score. The initial weights are inherited from models pre-trained for object recognition on ImageNet~\cite{deng2009imagenet}, expect for the last layer that is initialized by the method of He \textit{et al.} ~\cite{He2015Delving}. The two parameters $T$ and $\sigma$ that govern the reliability of PDIP generation are set to $0.95$ and $0.3$, respectively. During both training and testing, the short side of the input image is rescaled to $256$, from which a $224\times 224 \times 3$ sub-image is randomly cropped. 
 The training is carried out by optimizing the cross entropy function using Adam~\cite{kingma2014adam} with an $\ell_2$ penalty multiplier of $10^{-4}$  and a batch size of $64$. The learning rates for the pre-trained DNN layers and the last layer are set to $10^{-5}$ and $10^{-4}$, respectively.
 After each epoch, we decay the learning rates linearly by a factor of $0.95$.  Training takes approximately one day on an Intel E5-2699 2.2GHz CPU and an NVIDIA Tesla V100 GPU. Our model takes $450$ ms and $20$ ms to process an image of size $224\times224\times3$ on CPU and GPU, respectively. To facilitate research in I$^2$PA, we make the PyTorch implementation of our model and the large-scale image database publicly available at  \url{https://github.com/dingkeyan93/intrinsic-image-popularity}.

%---------------------Quantitative analysis----------------------------

\subsection{Quantitative evaluation}\label{subsec:quan}

\noindent\textbf{Main results on Instagram.} 
We adopt pairwise accuracy as the quantitative measure, which is defined as the percentage of correctly ranked pairs. From the $2.5$ million of PDIPs in the proposed database, we randomly choose $50,000$ pairs for validation, $50,000$ pairs for testing, and leave the rest for training. The weights that achieve the highest pairwise accuracy on the validation set are used for testing.

As a relatively new problem, it is difficult to find computational models specifically for I$^2$PA in the literature. We try our best to compare our model with four most relevant and state-of-the-art methods, whose implementations are publicly available for testing only. These are 
Khosla14~\cite{khosla2014makes}, Hessel17~\cite{hessel2017cats}, Virality Detection~\cite{virality}, and LikelyAI~\cite{likelyai}. Khosla14 makes one of the first attempts to predict absolute image popularity. It also provides an API to assess intrinsic image popularity. Hessel17 is a multi-modal content popularity predictor based on Reddit. Six category-specific models are trained, and the one (for the \textit{pics} category) that achieves the highest pairwise accuracy  on our test set is used for comparison. Virality Detection and LikelyAI are two commercialized products, aiming to predict image popularity in a variety of practical scenarios (with or without social and textual contexts). %To the best of our knowledge, LikelyAI is the only Instagram-based model, whose implementation is publicly available.
\begin{figure}[t]
\centering
  \includegraphics[width=0.85\linewidth]{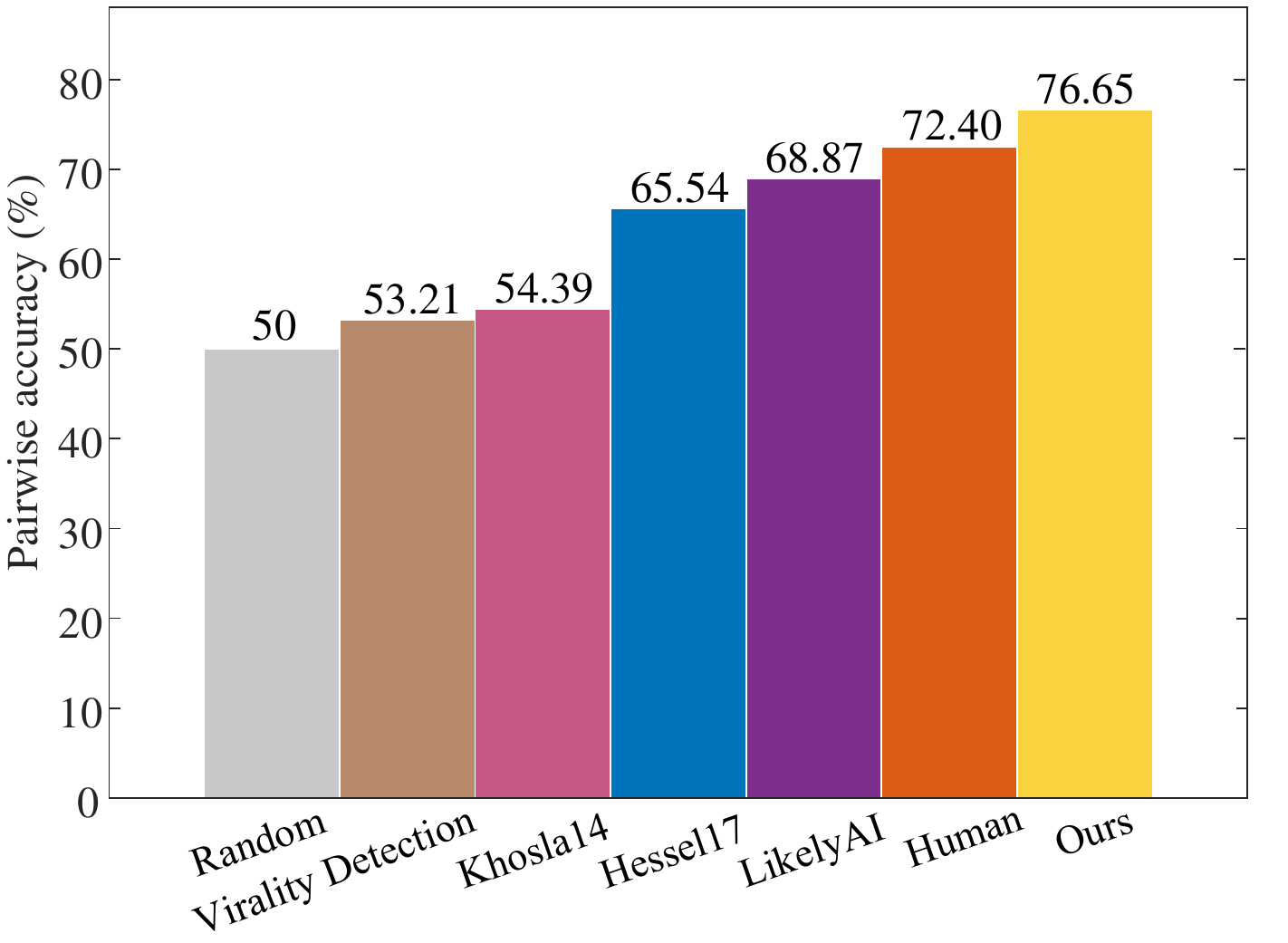} 
  \caption{Pairwise accuracy on the test set consisting of 50,000 PDIPs. Note that the human-level performance is measured on 1,000 PDIPs randomly sampled from the test set due to the high cost of the psychophysical experiment.}
  \label{fig:res}
%   \vspace{-3mm}
\end{figure}

Fig.~\ref{fig:res} shows the results, where we see that our model achieves the best performance with a pairwise accuracy of $76.65\%$. Khosla14 and Virality Detection marginally outperform the random guess baseline, which may be due to the distribution mismatch between training (Flickr) and testing (Instagram) images. Specifically, Instagram is a community conductive to self-expression, while Flickr focuses more on photographs of high visual quality and aesthetics. Hessel17 suffers from the similar issue, whose training data are crawled from Reddit rather than Instagram. Trained on Instagram images, LikelyAI performs slightly better than Hessel17, but is inferior to our model by a large margin. We believe this performance improvement arises because the PDIPs used for training contain reliable information about intrinsic image popularity, and our end-to-end optimized model is able to capture the features and attributes of images that are highly relevant to intrinsic image popularity. Our results also suggest that a fine-grained treatment of I$^2$PA based on different social platforms may be needed to combat data distribution mismatch.
 
%  Reddit and Flickr datasets
\vspace{2mm}
\noindent\textbf{Generalizability on Reddit and Flickr.}
To probe the generalizability of our model trained on Instagram, we test it on two other social platforms - Reddit and Flickr. The Reddit database~\cite{hessel2017cats} contains over $100,000$ pairs of popular and unpopular images categorized by six sub-datasets. We choose the largest sub-dataset \textit{pics} due to its diverse content variations. Our model achieves a pairwise accuracy of $58.9\%$ (on $44,343$ pairs), and is slightly worse than Hessel17~\cite{hessel2017cats} ($60.0\%$), which is trained on the same Reddit database.

Next, we test our model on images from Flickr. Due to the lack of intrinsic image popularity databases on Flickr, we decide to build a small one for testing. Specifically, we choose the social media headline prediction challenge database~\cite{Wu2017DTCN} as the starting point. The database contains over $340,000$ posts from over $80,000$ users. For simplicity, we select the most popular $50,000$ images and the most unpopular $50,000$ images according to the normalized number of views, and pair them randomly. When considering the visual content only, our model achieves a pairwise accuracy of $63.3\%$, and is slightly better than $62.4\%$ of Khosla14~\cite{khosla2014makes}, which is trained on the same Flickr database.
%In summary, we empirically show that the proposed model trained on millions of Instagram PDIPs exhibits reasonable generalizability to Reddit and Flickr. 

As previously discussed, the performance drop of our Instagram-based model on Reddit and Flickr may be because the dataset distributions are different. In addition,  without reducing the effect of non-visual factors, the test image pairs are much noisier.

\begin{table}[t]
  \centering
  \caption{Pairwise accuracy as a function of simulated label noise level}
  \setlength{\tabcolsep}{2.7mm}{
    \begin{tabular}{c|ccccc}
    \toprule
    Noise  & $0\%$ & $10\%$ & $20\%$ & $30\%$ & $40\%$ \\ 
    \hline
    Accuracy (\%) & $76.65$ & $75.61$ & $74.35$ & $73.10$ & $68.55$ \\
  \bottomrule
    \end{tabular}
    }
  \label{tab:acc_noise}
\end{table}
% \vspace{-3mm}

\begin{table}[t]
  \centering
  \caption{Pairwise accuracy as a function of network architecture}
    \begin{tabular}{c|cccc}
    \toprule
    DNN & 
    AlexNet &
    VGGNet & 
    ResNet-50 & 
    ResNet-101 \\
    \hline
    Accuracy (\%)  & $73.22$ & $76.15$ & $76.65$ & $76.87$ \\
    \bottomrule
    \end{tabular}
  \label{tab:acc_dnn}
\end{table}

% \begin{table*}[t]
%   \centering
%   \caption{Pairwise accuracy as a function of noise levels and DNN architectures. The default setting is highlighted in bold }
%     \begin{tabular}{c|cccccc}
%     \toprule
%     Noise  & $\mathbf{0}$ & $10\%$ & $20\%$ & $30\%$ & $40\%$ & $50\%$ \\ 
%     \hline
%     Accuracy & $76.65$ & $75.61$ & $74.35$ & $73.10$ & $68.55$ & $52.12$\\
%     \midrule
%     DNN & SqueezeNet~\cite{iandola2016squeezenet} & AlexNet~\cite{krizhevsky2012imagenet} & 
%     VGGNet~\cite{simonyan2014very} & 
%     \textbf{ResNet-50}~\cite{he2016deep} & 
%     DenseNet~\cite{huang2017densely} & 
%     ResNet-101~\cite{he2016deep}\\
%     \hline
%     Accuracy  & $72.98$ & $73.22$ & $76.15$ & $76.65$ & $76.71$ & $76.87$ \\
%     \bottomrule
%     \end{tabular}
%   \label{tab:accuracy}
% \end{table*}

%  ---------score distribution-----------
\begin{figure}[t]
\centering
  \includegraphics[width=0.8\linewidth]{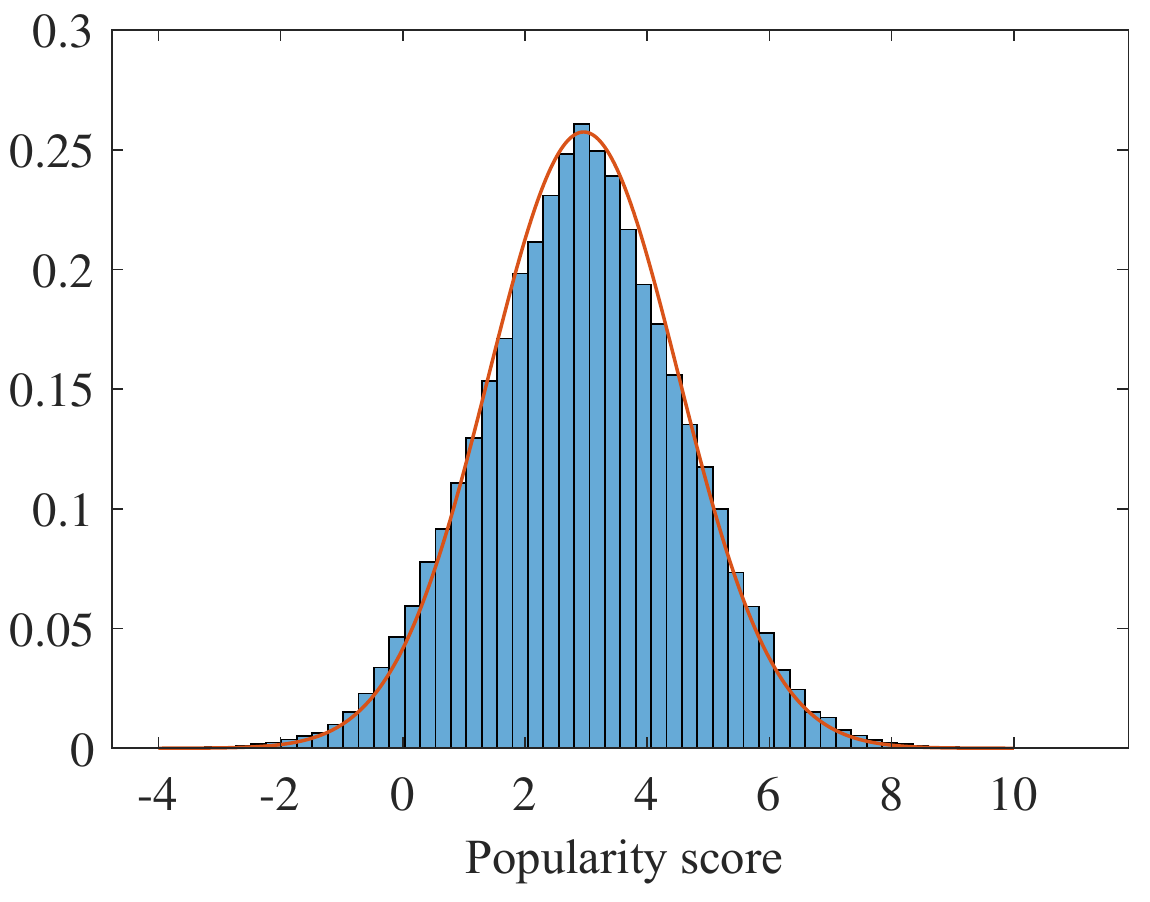}
  \caption{The normalized histogram of the predicted intrinsic popularity scores by our model on the test set with 50,000 PDIPs. Red line: fitted Gaussian curve.}
  \label{fig:histo}
\end{figure}

%  -----------5 levels image sets----------
\begin{figure*}[t]
  \centering
    \subfloat[]{\includegraphics[height=0.167\linewidth]{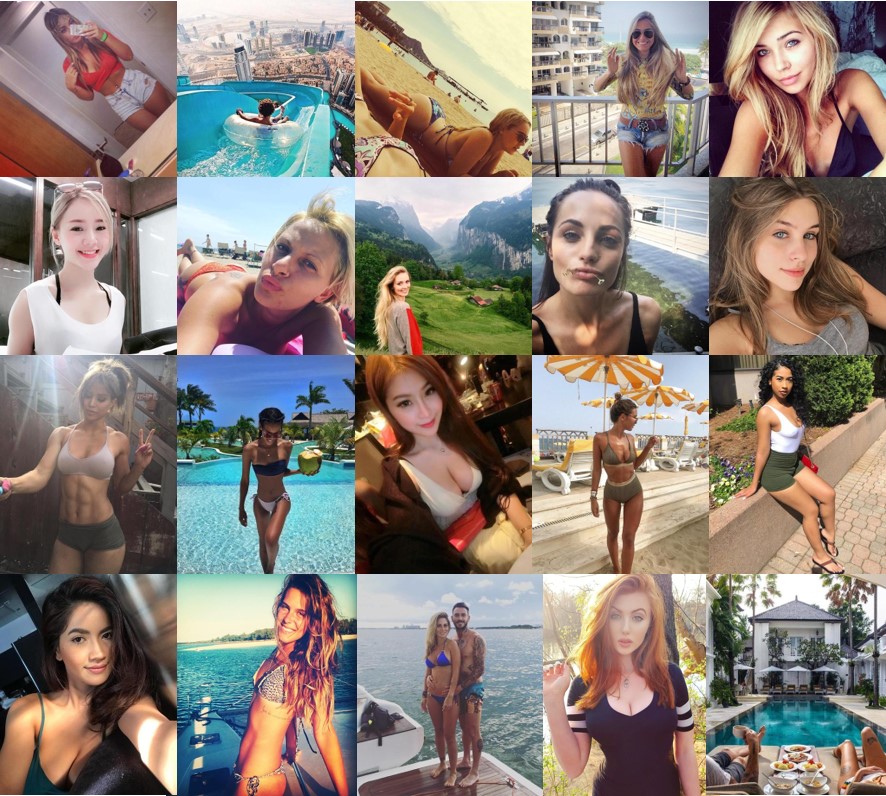}}\hskip.29em
    \subfloat[]{\includegraphics[height=0.167\linewidth]{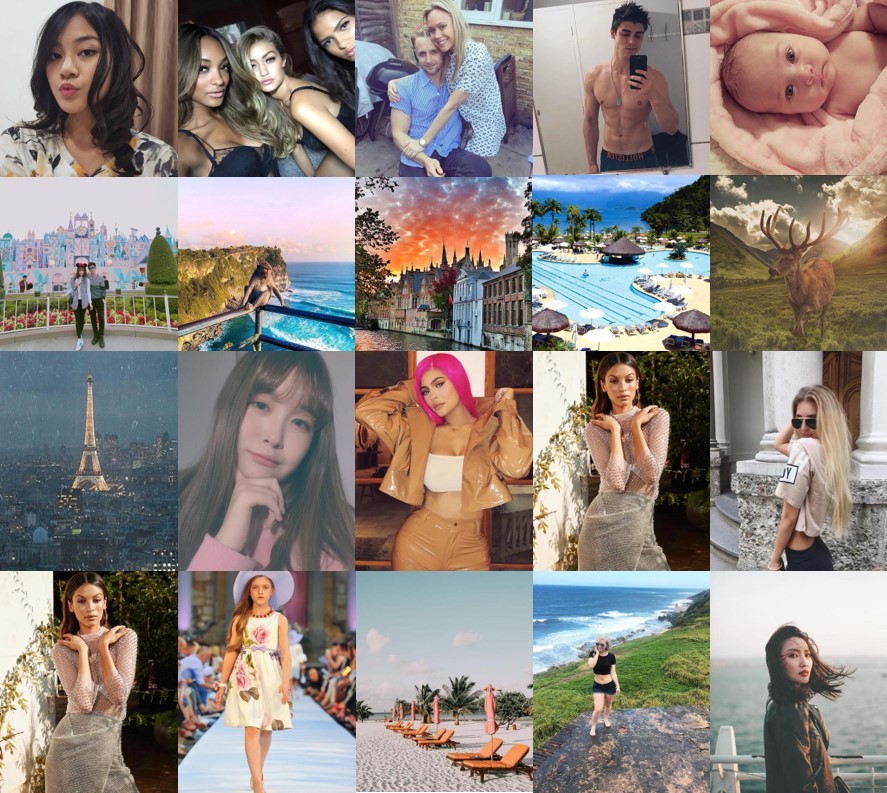}}\hskip.295em
    \subfloat[]{\includegraphics[height=0.167\linewidth]{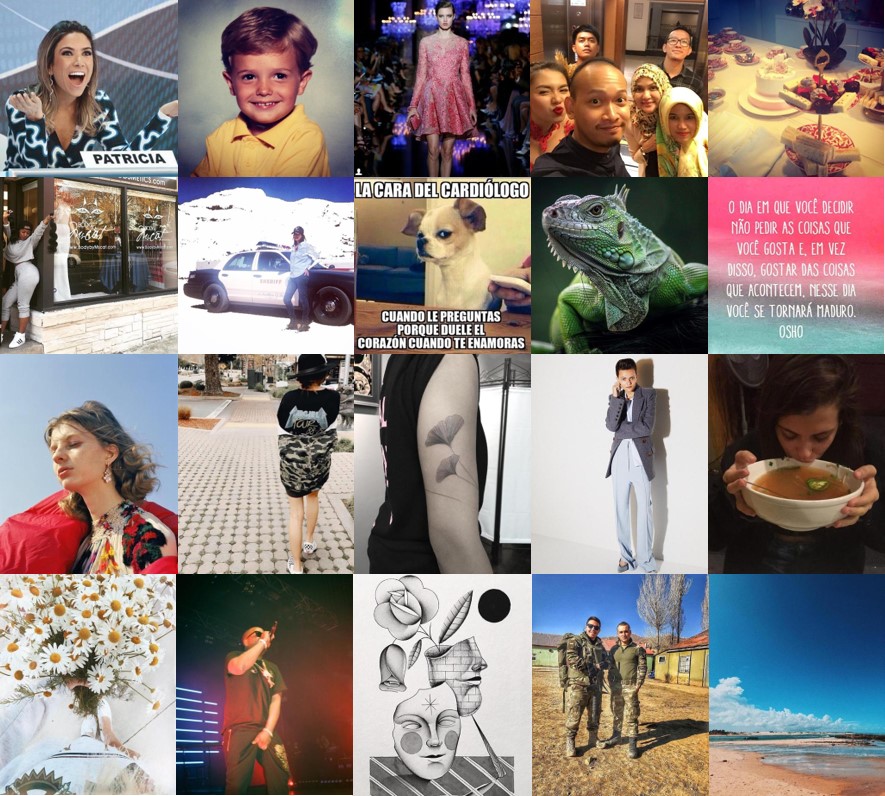}}\hskip.295em
    \subfloat[]{\includegraphics[height=0.167\linewidth]{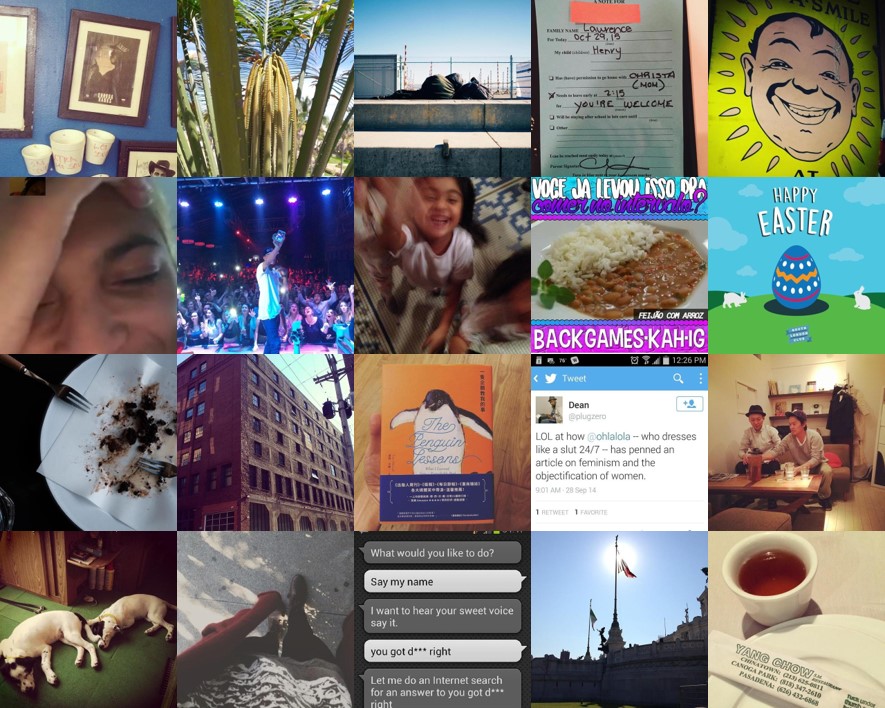}}\hskip.295em
    \subfloat[]{\includegraphics[height=0.167\linewidth]{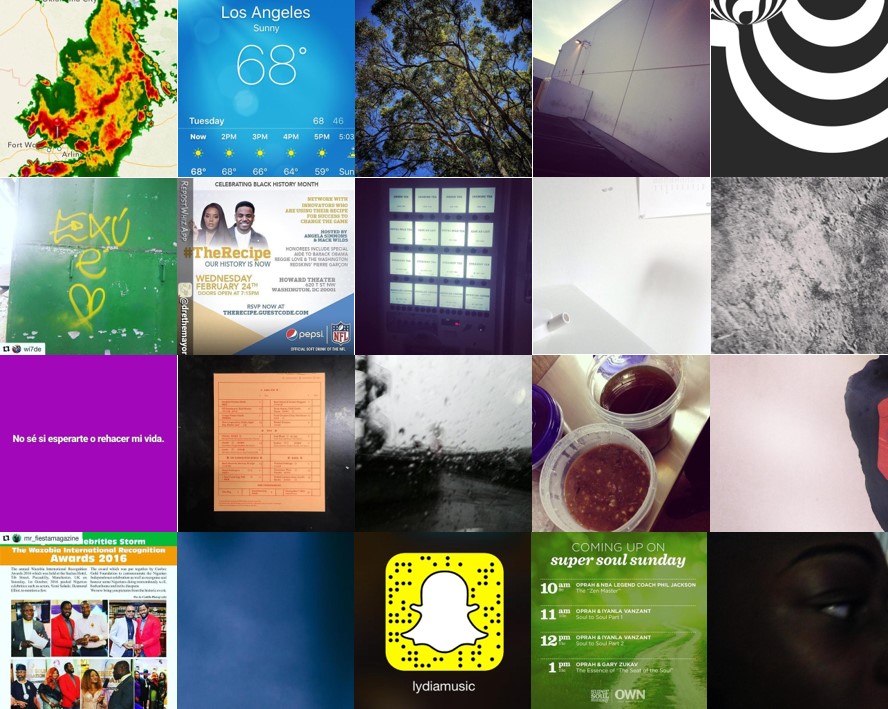}}
  \caption{Images with different predicted intrinsic popularity levels. Some images are resized without keeping aspect ratios for neat display. (a) Excellent. (b) Good. (c) Fair. (d) Bad. (e) Poor.}
  \label{fig:range}
\end{figure*}

% \begin{figure*}[t]
%   \centering
%     \subfloat[]{\includegraphics[height=0.29\linewidth]{fig/range_1.jpg}}\hskip.5em
%     \subfloat[]{\includegraphics[height=0.29\linewidth]{fig/range_2.jpg}}\hskip.5em
%     \subfloat[]{\includegraphics[height=0.29\linewidth]{fig/range_3.jpg}} \\
%     \subfloat[]{\includegraphics[height=0.29\linewidth]{fig/range_4.jpg}}\hskip.5em
%     \subfloat[]{\includegraphics[height=0.29\linewidth]{fig/range_5.jpg}}
%   \caption{Images with different predicted intrinsic popularity levels. Some images are resized without keeping the aspect ratio for neat display. (a) Excellent. (b) Good. (c) Fair. (d) Bad. (e) Poor.}
%   \label{fig:range}
% \end{figure*}

%  -----------activation heat map-----------
\begin{figure*}[htbp]
\centering
  \includegraphics[width=1\linewidth]{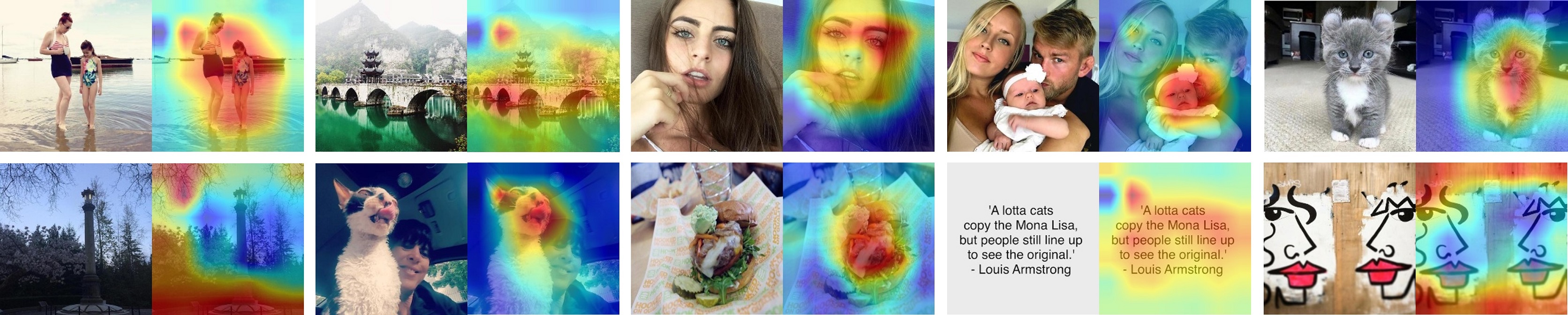}
  \caption{Heatmaps of sample images generated by Grad-CAM~\cite{selvaraju2017grad}. A warmer region contributes more to I$^2$PA. First row: images of high predicted popularity. Second row: images of low predicted popularity.}
  \label{fig:heatmap}
\end{figure*}

\vspace{2mm}
\noindent\textbf{Ablation study.} We provide a baseline for our model - a DNN trained for absolute IPA and used to assess intrinsic image popularity. Specifically, we first summarize the visual content by a scalar using the same ResNet-50~\cite{he2016deep}, and combine the image-based score with six non-visual features: the number of followers, followings, posts, hashtags, @ signs, and the length of caption. The seven-dimensional feature vector is fed to a fully connected neural network with a $7$-$256$-$128$-$64$-$1$ structure to predict the number of likes. We train the entire model end-to-end on $4.8$ million images used to generate training PDIPs. We adopt the mean squared error (MSE) as the loss function. The training procedure is the same as in Section~\ref{subsec:id}. The optimized model predicts absolute image popularity reasonably, as evidenced by a Pearson correlation of $0.83$ on the training set and $0.80$ on the test set, respectively. However, when tested on the $50,000$ PDIPs, this baseline model only achieves a pairwise accuracy of $66.5\%$. The performance drop may be because strong features (\eg, the number of followers) tend to dominate the learning, leaving less room for exploiting the visual content.

We analyze the impact of $T$ and $\sigma$ in PDIP generation on the final performance. However, due to the limited computation and storage resources, we perform a similar but simulated label noise experiment. In particular, we select a percentage $q\in \{10\%, 20\%, 30\%, 40\%\}$ to randomly flip the binary labels of PDIPs in the training set and retrain the models, whose test results are listed in Table~\ref{tab:acc_noise}. It is clear that the training is robust to label noise. When the label noise level is $30\%$, our model is still competitive with humans. Although it seems impossible to eliminate the impact of non-visual factors in our current PDIP generation process, our experiment suggests that the induced label noise does not seem to hinder the learning of a robust model for I$^2$PA. In addition, we may relax the constraints of PDIP generation (\ie, decrease $T$ or increase $\sigma$) to obtain pairs with more diverse content.

% 3 rows pairs with score and likes
% \begin{figure*}[t]
% \centering
%   \includegraphics[width=1\linewidth]{fig/pairs_score}
%   \caption{The intrinsic popularity predictions / the number of likes of representative PDIPs. Our model ranks the PDIPs in the first two rows correctly, while tends to make mistakes in the last row.}
%   \label{fig:ps}
%   \vspace{-3mm}
% \end{figure*}

\begin{table*}[t]
  \centering
  \caption{Pairwise accuracy of different strategies and factors. Majority: performance obtained by majority vote. Single: performance obtained by averaging individual subjects. $G_I$: subjects with Instagram experience. $G_{II}$: subjects with little Instagram experience. $G_A$: subjects spending more than three hours on social media per day. $G_B$: subjects spending less than three hours on social media per day}
    \begin{tabular}{l|cc|cc|cc|cc}
    \toprule
    Accuracy (\%) & Majority & Single & Female & Male & $G_I$ & $G_{II}$ & $G_A$ & $G_B$  \\ 
    \hline
    Mean & $72.4$ & $66.6$ & $67.0$ & $66.8$ & $68.5$ & $65.1$ & $67.1$ & $66.4$ \\
    Std & --- & $4.4$ & $4.2$ & $4.5$ & $3.7$ & $4.5$ & $3.8$ & $4.1$\\
    \bottomrule
    \end{tabular}
  \label{tab:results}
\end{table*}

We also investigate the impact of different DNN architectures on pairwise accuracy, including AlexNet~\cite{krizhevsky2012imagenet}, VGGNet~\cite{simonyan2014very}, ResNet-50~\cite{he2016deep} (default), and ResNet-101~\cite{he2016deep}. From the results in Table~\ref{tab:acc_dnn}, we find that there is still room for improvement on top of ResNet-50 if deeper and more advanced networks are adopted.

%-----------------Qualitative analysis------------------------

\subsection{Qualitative analysis}\label{subsec:qual}
We provide qualitative analysis of I$^2$PA from three perspectives: global image content (Fig.~\ref{fig:range}), local image content (Fig.~\ref{fig:heatmap}), and comparison with human data (Fig.~\ref{fig:4part}). It should be noted that these results would become less obvious if we predict absolute image popularity and do not single out the contribution of image content.

We first exam the histogram of predicted intrinsic popularity scores by our model on the test set, which provides a good coverage of representative Instagram images. The histogram can be well fitted by a Gaussian distribution with mean $2.96$ and std $1.55$. A higher value indicates better intrinsic popularity (see Fig.~\ref{fig:histo}).

To better analyze how global image content affects I$^2$PA, we define five popularity levels - ``excellent'', ``good'', ``fair'', ``bad'', and ``poor'' that evenly cover the predicted score range. Fig.~\ref{fig:range} shows representative images of each level. We find that images in the excellent level are often beautiful and attractive people, which is in close agreement with  Park and Lee's conclusion~\cite{park2017private}. Images in the good level tend to be brilliant selfies and spectacular sceneries. The high-score selfies are often accompanied by beautiful faces, which is consistent with the result that photos with faces are $38\%$ more likely to receive likes~\cite{bakhshi2014faces}.
Images in the fair level look ordinary and forgettable, whose common characteristics are difficult to summarize because of the content diversity. Images in the bad level are less prominent, and may lack interesting and distinguishable features. Images in the poor level  mainly consist of empty backgrounds with few salient objects.

We also investigate how local image content contributes to I$^2$PA. Specifically, we generate the heatmaps of sample images by Grad-CAM~\cite{selvaraju2017grad, bielski2018pay}, a visual explanation for deep networks via gradient-based localization. A warmer region in the heatmap indicates that it plays a more important role in I$^2$PA. From the first row of Fig.~\ref{fig:heatmap}, we find that several elements such as fine architectures (the second image), pretty faces (the third image), lovely kids (the fourth image), and cute animals (the fifth image) are often activated, leading to high popularity predictions. By contrast, for images in the second row, poor quality regions (the first image), unsightly expressions (the second image), textual descriptions (the fourth image), and empty backgrounds (the fifth image) tend to dominate, leading to low popularity predictions.

%---------------------Psychophysical experiment--------------------------

\begin{figure}[t]
\centering
  \includegraphics[width=1\linewidth]{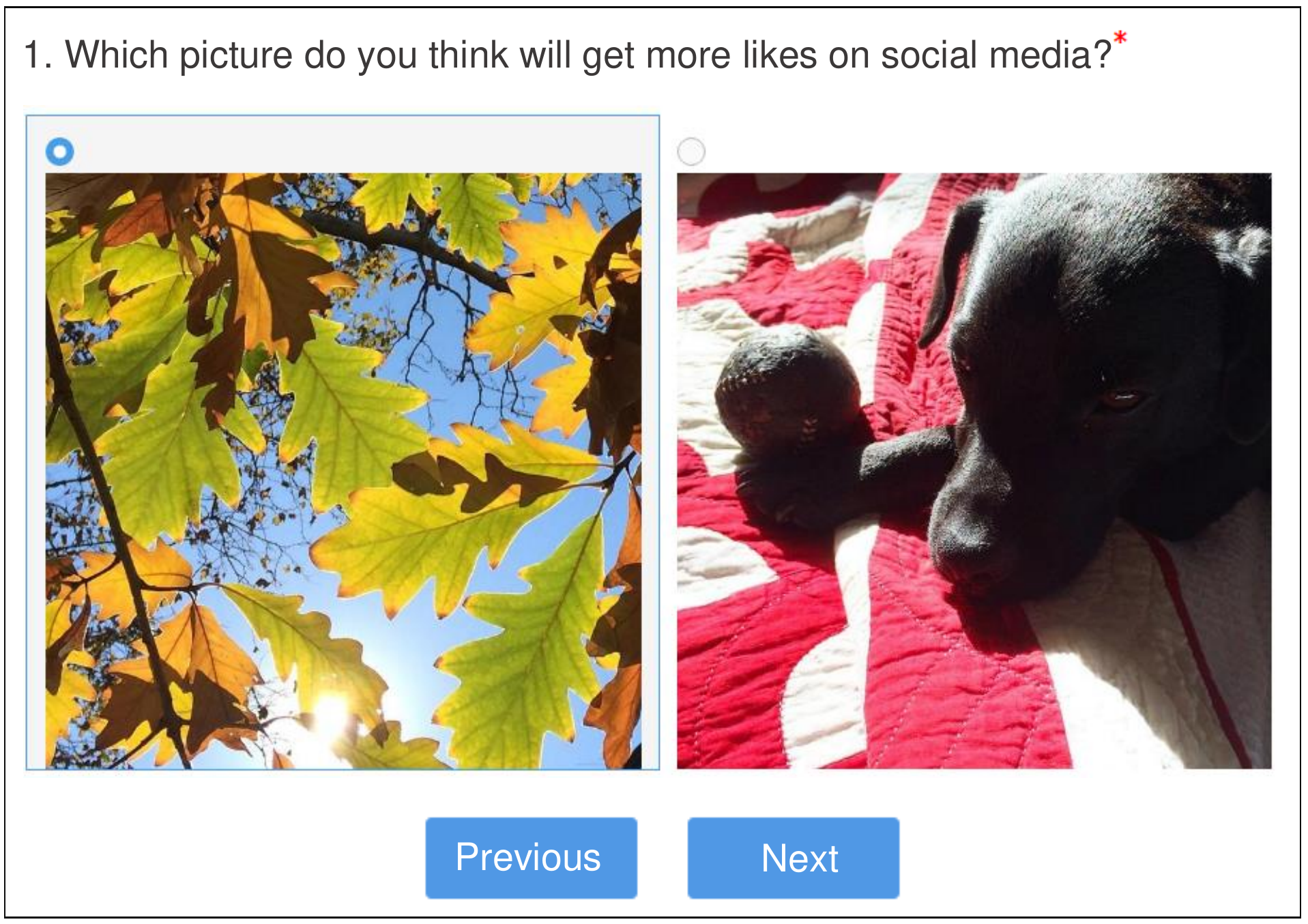}
  \caption{Web-based user interface for the psychophysical experiment.}
  \label{fig:web}
%   \vspace{-3mm}
\end{figure}

\subsection{Psychophysical experiment} 

To understand human behavior in I$^2$PA and to make the results more interpretable, we conduct a psychophysical experiment. Specifically, we randomly select $1,000$ PDIPs from the test set, and invite $30$ subjects to perform a two-alternative forced choice (2AFC) using a web-based platform (see Fig.~\ref{fig:web}). The subjects ($14$ females and $16$ males) are all college students of age $20$ to $30$, among whom $40\%$ have used Instagram before. Besides, $30\%$ subjects spend more than three hours on various social platforms every day. At the beginning of the experiment, eight training pairs are displayed to help the subjects build the concept of intrinsic image popularity. 
After that, they are free to make decisions based on their own understanding of image popularity. To reduce the fatigue effect, the experiment is divided into four sessions, each of which is limited to a maximum of $30$ minutes. The subjects are encouraged to participant in multiple sessions. %The spatial and temporal orders of PDIPs are randomized. In addition, we set three PDIPs in each session that have clear popularity discriminability as sanity check pairs. Anyone who fails in more than one pair will be removed. 
In the end, the subjects are allowed to review and compare their choices with the ground truths. 

%After removing one subject that fails in the sanity check, we obtain $17$, $17$, $15$, and $15$ valid responses in Session I, Session II, Session III, and Session IV, respectively. 
Table~\ref{tab:results} lists the subjective results, where we see that the majority vote strategy significantly outperforms individual subjects, reflecting the difficulty of this task for a single observer. 
We next analyze the influence of the gender, the Instagram experience, and the online social time per day on I$^2$PA. %Specifically, we first separate the subjects according to their gender and
From table~\ref{tab:results}, we find that 1) both male and female subjects tend to perform at a similar level; 2)  subjects with Instagram experience (denoted by $G_I$) perform statistically better (based on t-statistics~\cite{montgomery2007applied}) than those with little knowledge about Instagram (denoted by $G_{II}$); 3) subjects who spend more than three hours on social media per day (denoted by $G_A$) perform statistically better than those with less online social experience (denoted by $G_B$).

\begin{figure*}
\centering
  \includegraphics[width=1\linewidth]{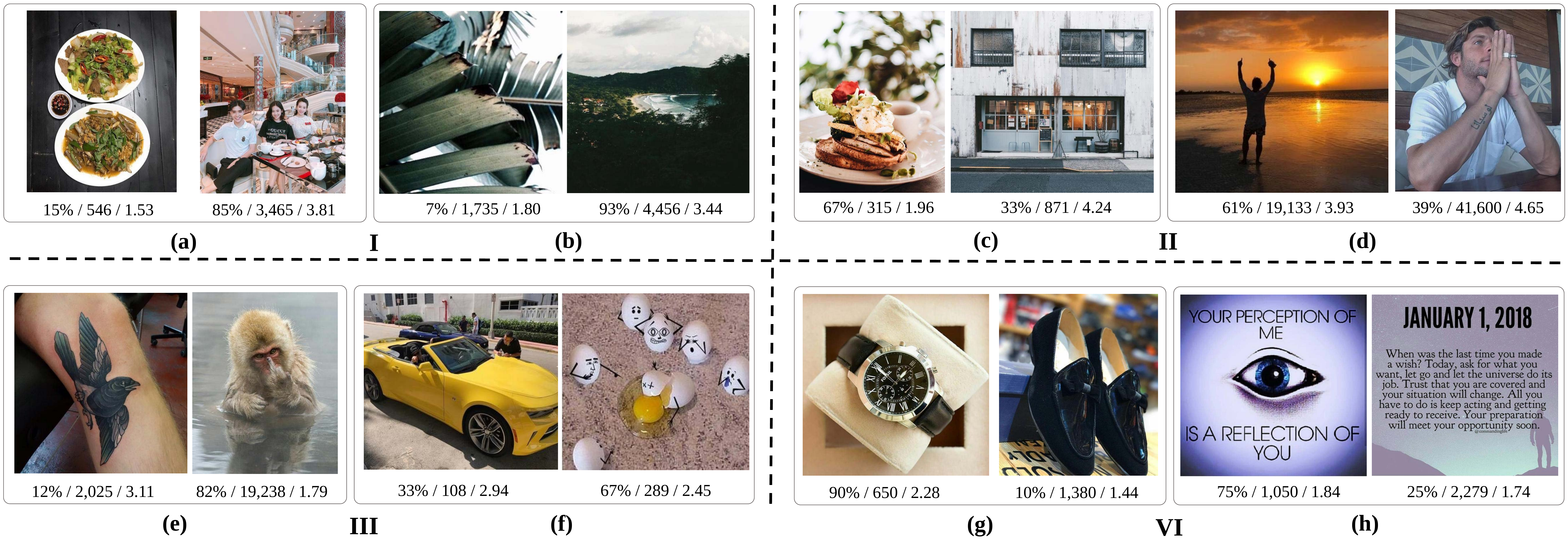}
  \caption{Eight representative PDIPs, within which the right image is intrinsically more popular than the left one. Below each image show the subject percentage in favor of the image, the number of received likes, and the popularity score predicted by our model, respectively.}
  \label{fig:4part}
\end{figure*}

We further compare the subjects' choices against our model predictions and the ground truths using four types of pairs, as shown in Fig.~\ref{fig:4part}.
%To further investigate human behavior in I$^2$PA, we show the choices made by the subjects and the proposed computational model versus the ground truths (\ie, the number of likes). The pairs set are divided into four parts: I) Both the subjects and our model are consistent with the ground truths. II) Our model do agree with the ground truths, but the subjects do not. III) The subjects do agree with the ground truths, but our model do not. VI) Neither the subjects nor our model are consistent with the ground truths. We display two representative pairs in each part, as shown in Fig.~\ref{fig:4part}.
Pairs (a) and (b) in Part I have clear popularity discriminability based on different visual appearances, leading to easy predictions for both subjects and our model. Most PDIPs in the proposed database belong to this category. 

For the pair (c) in Part II, $67\%$ of the subjects predict the food image to receive more likes than the house image. However, we find that the user has posted too many food images, of which the followers may get tired. The more likes received by the house image indicates that the followers are more interested in viewing images with novel content. This contextual interactions among posts complicate I$^2$ PA because humans and our model do not get access to such information. For the pair (d), many subjects pay too much attention to image aesthetics (\ie, they think the left image is more beautiful), which often results in selection bias. On Instagram, brilliant selfies generally receive more likes, which has been successfully captured  by our model. 

The pairs in Part III are difficult for our model because extremely abstract attributes such as the peculiar gesture (the pair (e)) and the creative/funny content (the pair (f)) need to be parsed and transformed to the concept of popularity. By contrast, humans have a better understanding of these concepts, and are able to make consistent choices easily.
 
For the pair (g) in Part VI, nearly all subjects prefer the elegant watch than the ordinary shoes, and our model agrees on this point. However, the shoes image receives more likes. We conjecture that the shoes may convey a special meaning (\eg, as a memorable gift), of which the subjects in the psychophysical experiment are unaware. The number of likes may also be boosted by the Internet vendor for sales promotion. For the pair (h), the number of likes is mainly determined by the texts in the image, which are difficult for most subjects to comprehend due to the cultural differences. Our model also fails to understand the words, and tends to give text images low popularity scores.

%Based on the results of psychophysical experiment, we have four interesting observations. First, I$^2$PA is difficult for humans due to the limited cognitive capability of a single observer and the diversity of the image content on social media. Second, the majority vote strategy significantly improves the average performance of individual subjects. Third, the rich experience of online social media is helpful to make correct predictions, especially the experience of using Instagram where the test PDIPs come from. Fourth, most subjects tend to favor images, which contain beautiful people, exquisite objects, bright colors, and few texts.

% \begin{figure*}[t]
% \renewcommand\thefigure{S1}
% \centering
%   \includegraphics[width=1\linewidth]{fig/sup_train}
%   \caption{Eight training PDIPs with the number of received likes for the psychophysical experiment.}
%   \label{fig:train}
% \end{figure*}

% \begin{figure*}[t]
% \renewcommand\thefigure{S2}
%   \centering
%     \subfloat[]{\includegraphics[width=1\linewidth]{fig/sup_t1.jpg}}\\
%     \subfloat[]{\includegraphics[width=1\linewidth]{fig/sup_t2.jpg}}\\
%     \subfloat[]{\includegraphics[width=1\linewidth]{fig/sup_t3.jpg}}\\ \subfloat[]{\includegraphics[width=1\linewidth]{fig/sup_t4.jpg}}
%   \caption{Sanity check pairs in each session. The right image in each pair has many more likes than the left one. (a) Session I. (b) Session II. (c) Session III. (d) Session IV.}
%   \label{fig:quality}
% \end{figure*}

%=============================================================================

\section{Conclusion}
We have conducted a systematic study of I$^2$PA. The principle behind I$^2$PA is to predict image popularity based on the image content only, and the concept of PDIP is introduced to reliably infer intrinsic popularity. The first large-scale image database for I$^2$PA is established, and a DNN-based computational model is further proposed, which achieves human-level performance. In addition, we have carried out a psychophysical experiment to understand how humans tend to behave in this task.

\section*{Acknowledgements}
The authors would like to thank Dr. Zhuo (Jimmy) Wang for insightful discussions on the probabilistic formulation of I$^2$PA. 
%We introduced the concept of PDIP, from which intrinsic image popularity can be reliably inferred, and built the first large-scale database for I$^2$PA. We developed a DNN-based computational model for I$^2$PA, which achieves human-level performance. Moreover, we carried out a psychophysical experiment to understand how humans tend to behave in this task.

% \section{Acknowledgments}
% \section{Appendices}
%
% The next two lines define the bibliography style to be used, and the bibliography file.
\bibliographystyle{ACM-Reference-Format}
% \bibliography{reference}

% 
% If your work has an appendix, this is the place to put it.
% \appendix

\end{document}